\documentclass[conference]{IEEEtran}

\usepackage{amsmath,amssymb,amsfonts}
\usepackage{algorithmic}
\usepackage{graphicx}
\usepackage{textcomp}
\usepackage{xcolor}
\usepackage{siunitx}
\usepackage{graphicx}

\usepackage{xcolor}

\usepackage{amsmath}
\usepackage{amssymb}
\usepackage{bm}
\usepackage{nicefrac}

\usepackage{rotating, tabularx}
\usepackage{array}
\usepackage{booktabs}
\usepackage{multirow}
\usepackage{pdflscape}
\usepackage{afterpage}
\usepackage{makecell}
\usepackage{adjustbox}
\usepackage{graphicx}
\usepackage{threeparttablex}

\usepackage[utf8]{inputenc}
\usepackage{textcomp,marvosym}

\usepackage{hyperref}
\hypersetup{hidelinks}

\usepackage[english]{babel}
\usepackage{csquotes}
\usepackage[nohyperlinks,nolist]{acronym}
\acrodef{AI}[AI]{artificial intelligence}
\acrodef{ANN}[ANN]{artificial neural network}
\acrodef{AC}[AC]{accumulate}
\acrodef{BMI}[BMI]{brain-machine interface}
\acrodef{CST}[CST]{cortical spike train}
\acrodef{DNN}[DNN]{deep neural network}
\acrodef{SNN}[SNN]{spiking neural network}
\acrodef{CNN}[CNN]{convolutional neural network}
\acrodef{GPU}[GPU]{graphics processing unit}
\acrodef{CPU}[CPU]{central processing unit}
\acrodef{MAC}[MAC]{multiply-and-accumulate}
\acrodef{MUA}[MUA]{multi-unit activity}
\acrodef{M1}[M1]{primary motor cortex}
\acrodef{S1}[S1]{primary sensory cortex}
\acrodef{SGD}[SGD]{stochastic gradient descent}
\acrodef{LIF}[LIF]{leaky integrate-and-fire}
\acrodef{LI}[LI]{leaky integrator}
\acrodef{I1}[I1]{indy\_20160622\_01}
\acrodef{I2}[I2]{indy\_20160630\_01}
\acrodef{I3}[I3]{indy\_20170131\_02}
\acrodef{L1}[L1]{loco\_20170210\_03}
\acrodef{L2}[L2]{loco\_20170215\_02}
\acrodef{L3}[L3]{loco\_20170301\_05}
\acrodef{PSP}[PSP]{postsynaptic potential}
\acrodef{R²}[$R^2$]{coefficient of determination}
\acrodef{RSNN}[RSNN]{recurrent spiking neural network}
\acrodef{RMSE}[RMSE]{root mean squared error}
\acrodef{SHD}[SHD]{Spiking Heidelberg Digits}
\acrodef{ResNet}[ResNet]{residual network}
\acrodef{SG}[SG]{surrogate gradient}
\acrodef{CSNN}[CSNN]{convolutional spiking neural network}
\acrodef{ReLU}[ReLU]{rectified linear unit}
\acrodef{BPTT}[BPTT]{back-propagation through time}

\usepackage[
natbib=true,
style=ieee]{biblatex}
\title{Decoding finger velocity from cortical spike trains with recurrent spiking neural networks}

\author{
\IEEEauthorblockN{
Tengjun Liu\textsuperscript{1,2,3,*}, Julia Gygax\textsuperscript{3,4,*}, Julian Rossbroich\textsuperscript{3,4,*},\\ Yansong Chua\textsuperscript{5}, Shaomin Zhang\textsuperscript{1,2,\dag}, Friedemann Zenke\textsuperscript{3,4,\dag}}
\IEEEauthorblockN{\tiny{ }}
\IEEEauthorblockA{\textsuperscript{1}\textit{Qiushi Academy for Advanced Studies, Zhejiang University}, Hangzhou, China}
\IEEEauthorblockA{\textsuperscript{2}\textit{College of Biomedical Engineering and Instrument Science, Zhejiang University}, Hangzhou, China}
\IEEEauthorblockA{\textsuperscript{3} \textit{Friedrich Miescher Institute for Biomedical Research}, Basel, Switzerland}
\IEEEauthorblockA{\textsuperscript{4} \textit{Faculty of Science, University of Basel}, Basel, Switzerland}
\IEEEauthorblockA{\textsuperscript{5} \textit{China Nanhu Academy of Electronics and Information Technology (CNAEIT)}, Jiaxing, China}
\IEEEauthorblockN{\tiny{ }}
\IEEEauthorblockA{\textsuperscript{*} These authors contributed equally to this work.}
\IEEEauthorblockA{\textsuperscript{\dag} Corresponding authors: \href{mailto:shaomin@zju.edu.cn}{shaomin@zju.edu.cn}, \href{mailto:friedemann.zenke@fmi.ch}{friedemann.zenke@fmi.ch}}
}

\date{}

\begin{document}
\maketitle

\begin{abstract}
    Invasive cortical \acp{BMI} can significantly improve the life quality of motor-impaired patients.
    Nonetheless, externally mounted pedestals pose an infection risk, which calls for fully implanted systems.
    Such systems, however, must meet strict latency and energy constraints while providing reliable decoding performance.
    While \acp{RSNN} are ideally suited for ultra-low-power, low-latency processing on neuromorphic hardware, it is unclear whether they meet the above requirements.
    To address this question, we trained \acp{RSNN} to decode finger velocity from \acp{CST} of two macaque monkeys.
    First, we found that a large \ac{RSNN} model outperformed existing feed-forward \acp{SNN} and \acp{ANN} in terms of their decoding accuracy.
    We next developed a tiny \ac{RSNN} with a smaller memory footprint, low firing rates, and sparse connectivity.
    Despite its reduced computational requirements, the resulting model
    performed substantially better than existing \ac{SNN} and \ac{ANN} decoders.
    Our results thus demonstrate that \acp{RSNN} offer competitive \ac{CST} decoding performance under tight resource constraints and are promising candidates for fully implanted ultra-low-power \acp{BMI} with the potential to revolutionize patient care.
\end{abstract}


\acresetall

\section{Introduction}
\Acp{BMI} enable direct communication between biological neural networks and external devices \citep{chaudhary2016brain}.
Significant progress has been made in invasive \ac{BMI} technology, allowing volitional control of robotic arms \citep{hochberg2012reach, flesher2021brain} and text generation for communication \citep{pandarinath2017high, willett2021high, moses_neuroprosthesis_2021, Card2024-rv}.
Despite rapid improvements in decoding performance, a central challenge persists: the risk of wound infection
due to externally mounted pedestals.
Fully implanted \acp{BMI} offer a potential solution \citep{yik_neurobench_2024, taeckens_spiking_2024, liao2022energy}, but introduce new constraints on energy consumption and heat dissipation of the implants.
For instance, the American Association of Medical Instrumentation guidelines stipulate that chronically implanted medical devices must not increase tissue temperature by more than 1°C \citep{wolf2008thermal}.
Consequently, fully implanted \acp{BMI} require a computational substrate capable of delivering reliable and accurate decoding performance within strict latency and power limits.

\Acp{SNN} offer an attractive solution for fully implanted \acp{BMI}.
Their spike-based communication minimizes pre-processing requirements for \acp{CST} and makes them suitable for ultra-low power, low-latency neuromorphic hardware \citep{indiveri_neuromorphic_2011, donati_neuromorphic_2024}.
Several studies have explored the decoding abilities of feed-forward \acp{SNN} \citep{yik_neurobench_2024, taeckens_spiking_2024, liao2022energy, zhou_ieee_2024}, but these models achieved only moderate decoding performance on standard benchmarks, such as decoding finger velocity from monkeys performing a reaching task \citep{odoherty_nonhuman_2017}.
For instance, \citet{liao2022energy} developed an energy-efficient feed-forward \ac{SNN}, which achieved only moderate decoding accuracy (\ac{R²} value of 0.45 averaged across two \ac{CST} recording sessions).
The community-driven NeuroBench project \citep{yik_neurobench_2024} proposed additional \ac{SNN} models for \ac{CST} decoding, termed \textit{SNN} and \textit{SNN\_Flat}, achieving average \ac{R²} values of $0.58$ and $0.63$, respectively.
However, the increased decoding accuracy of the \textit{SNN\_Flat} model comes at a substantial energy cost.
Consequently, existing feed-forward \ac{SNN} models fail to simultaneously meet decoding performance and energy efficiency requirements.

\Acp{RSNN}, which go beyond simple feed-forward architectures, remain largely unexplored for fully implanted \acp{BMI} applications.
Here, we address this gap by investigating the decoding performance and energy efficiency of \acp{RSNN} on finger velocity decoding from \acp{CST} in monkeys \citep{odoherty_nonhuman_2017}.

\section{Methods}
To study the decoding performance of \acp{RSNN}, we iteratively developed two distinct model architectures:
First, we explored the maximum decoding performance achievable through end-to-end training of \acp{RSNN}, which we refer to as ``bigRSNN''.
Second, to balance decoding performance and energy efficiency for fully implanted \ac{BMI} applications, we developed a lightweight ``tinyRSNN'', reducing the number of model parameters by several orders of magnitude.

\subsection{Network architectures}
Both models consist of a spiking input layer matching the number of electrode channels in the \ac{CST} recording, followed by a single recurrent layer of standard \ac{LIF} neurons, and a readout layer of non-spiking \ac{LI} neurons (Fig.~\ref{fig:setup}A).
All \ac{LIF} and \ac{LI} units in the network feature synaptic and membrane dynamics with learnable, unit-specific synaptic and membrane time constants \citep{perez-nieves_neural_2021}.
During training, we optimized the membrane potential of the readout units to match the measured monkey finger velocities.
The two network models primarily differ in the size of their hidden and readout layers, with tinyRSNN additionally incorporating techniques that encourage activity and connection sparsity (see Section~\ref{sec:energy-saving}).

To explore the upper limit of decoding capabilities, we designed bigRSNN with 1024~units in the hidden layer.
Mimicking ensemble methods that have been proven successful in machine learning \citep{Rokach2005}, we incorporated a readout layer with five readout heads.
Each head consists of two non-spiking \ac{LI} neurons corresponding to finger velocities along the X and Y axes, respectively.
These readout heads were allowed to have different synaptic and membrane dynamics.
The final predicted finger velocities were determined by averaging the predictions across all readout heads.

To better accommodate the resource-constrained setting of realistic \ac{BMI} applications, we designed tinyRSNN with a limited hidden layer size of 64 recurrently connected \ac{LIF} neurons.
In contrast to the multi-head readout strategy of bigRSNN, tinyRSNN employs a simpler readout layer with only two \ac{LI} units, one each for X and Y directions.

\subsection{Data pre-processing}
\begin{figure}[tb]
    \centering
    \includegraphics[width=\columnwidth]{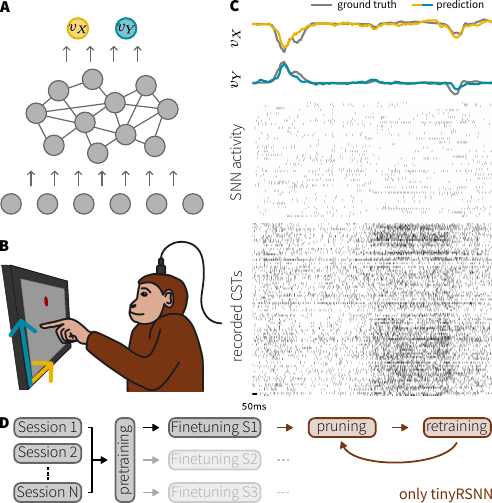}
    \caption{\textbf{Setup and example network activity.}
        \textbf{A)}~Schematic of the \ac{RSNN} network architecture.
        \textbf{B)}~The publicly available dataset consists of \acp{CST} recorded from two macaque monkeys performing a self-paced reaching task using chronically implanted electrodes.
        Recording sites were either in the primary motor cortex (M1) (\textit{Indy)} or in M1 and the primary sensory cortex (S1) (\textit{Loco})~\citep{odoherty_nonhuman_2017}.
        \textbf{C)}~Example network activity for tinyRSNN.
        Bottom: Spike raster of recorded \acp{CST} which serves as input to the model.
        Middle: Spike raster of the recurrent hidden layer activity.
        Top: Membrane potentials of the readout units (colored) and the ground-truth finger velocities (gray).
        \textbf{D)}~Schematic of the training curriculum.
        For each monkey, we pre-trained on all available sessions, with subsequent session-wise fine-tuning.
        For tinyRSNN, we further added iterative pruning at the end of the curriculum.
    }
    \label{fig:setup}
\end{figure}

Both models were trained on \ac{MUA} from \ac{CST} recordings obtained over multiple sessions from two macaque monkeys, Indy and Loco \citep{odoherty_nonhuman_2017, zhou_ieee_2024} (Fig.~\ref{fig:setup}B).
For each session, we split the data into 65\% training, 10\% validation, and 25\% test sets.
We discretized input spikes into 4\,ms long bins, each representing one time step for the model.
To enable mini-batch training, we divided the training and validation data into overlapping two-second samples (500 time steps per sample).
In contrast, we presented the test data to the models continuously, time step by time step as a single long sample, mimicking real-time processing conditions.

\subsection{Training}
The training process for bigRSNN and tinyRSNN followed the same structured approach:
We initialized models in the fluctuation-driven regime \citep{rossbroich_fluctuation-driven_2022}.
For each monkey, we first pre-trained the models for 100 epochs on concatenated and shuffled data samples from all available sessions \citep{odoherty_nonhuman_2017}.
After pre-training, we independently fine-tuned the models for 200 epochs on each of three sessions per monkey\footnote{Sessions: I1:~indy\_20160622\_01, I2:~indy\_20160630\_01, I3:~indy\_20170131\_02, L1:~loco\_20170210\_03, L2:~loco\_20170215\_02, and L3:~ loco\_20170301\_05.} \citep{zhou_ieee_2024} (cf. Fig.~\ref{fig:setup}D).
For each session, we trained five models with different parameter initializations.
All \acp{SNN} were trained using surrogate gradient descent \citep{neftci_surrogate_2019} to minimize the root mean squared error between predicted and target finger velocities.
We used the SMORMS3 optimizer \citep{funk_rmsprop_2015-1, rossbroich_fluctuation-driven_2022} with a cosine learning rate schedule to accelerate learning.
To avoid silent neurons during training, we added a homeostatic activity regularization term to the loss function \citep{rossbroich_fluctuation-driven_2022}.
Finally, we applied dropout with $p=0.3$ to the input and hidden layers to mitigate overfitting.
Detailed hyperparameters and training settings for both models can be found in the configuration files in our GitHub repository\footnote{Code repository: \url{https://github.com/fmi-basel/neural-decoding-RSNN}}.
All simulations were run in Python version 3.10.12 using custom code based on the Stork \ac{SNN} simulator \citep{rossbroich_fluctuation-driven_2022} and PyTorch \citep{paszke_pytorch_2019}.

\subsection{Activity regularization and pruning}
\label{sec:energy-saving}
We further optimized the tinyRSNN model for energy efficiency through three approaches:
First, we encouraged sparse neuronal activity by incorporating an activity regularization term to the loss function, which imposes an upper bound on the average firing rate of the hidden layer \citep{rossbroich_fluctuation-driven_2022}.
Second, we implemented an iterative pruning strategy for all synaptic weights to reduce the number of parameters and synaptic operations.
Following fine-tuning, we initially pruned the 40\% of the weights with the smallest magnitude and trained the remaining weights for an additional 100 epochs.
We then iteratively pruned an additional 10\% of weights and fine-tuned, as long as the \ac{R²} value stayed above a predefined threshold, i.e. 2\% below the original accuracy before pruning.
Once this threshold was reached, we reduced the pruning rate to 5\% and continued until the \ac{R²} value fell below the threshold again (Fig.~\ref{fig:setup}D).
Finally, to further reduce the memory footprint of the network, we converted all model parameters and buffers to half precision after training.

\section{Results}
We assessed the trained models' decoding performance and computational complexity using the following metrics as suggested by the NeuroBench project \citep{yik_neurobench_2024}:
For decoding performance, we calculated the \ac{R²} between predicted and observed finger velocities on the test dataset.
To evaluate computational cost, we measured the total memory footprint of the networks in bytes, their activation and connection sparsity, and the number of synaptic operations used by the models.
The latter was determined using the total number of operations during inference (Dense) and the total number of effective \ac{AC} and \ac{MAC} operations.
Notably, as both models are pure \ac{SNN} models without normalization layers, they perform no effective \ac{MAC} operations.
If not stated otherwise, we report averages across five different random initializations $\pm$ standard deviation.

\subsection{Decoding performance}
\begin{figure}[tb]
    \centering
    \includegraphics[width=\columnwidth]{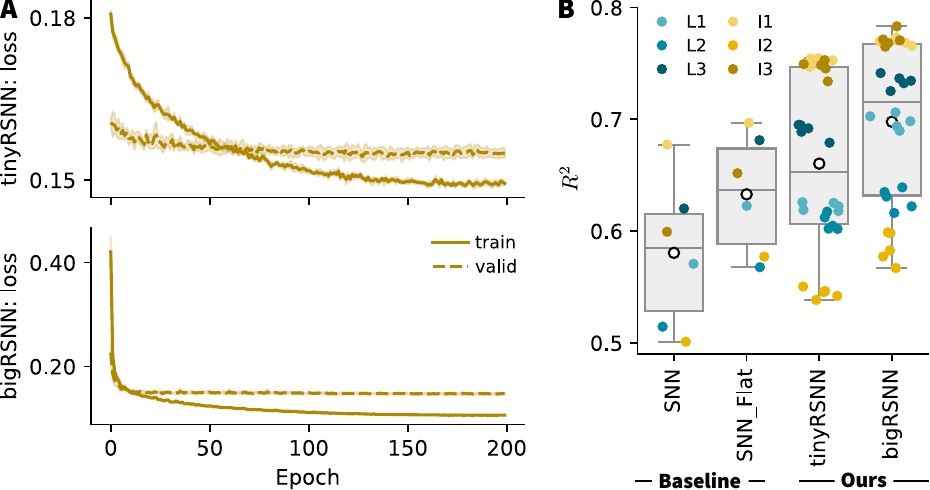}
    \caption{ \textbf{Learning and decoding performance of the \ac{RSNN} models.}
        \textbf{A)}~Learning curves for the tinyRSNN (top) and the bigRSNN (bottom) for session I3.
        \textbf{B)}~\ac{R²} values for ours and the baseline models (\textit{SNN}, \textit{SNN\_Flat} \citep{yik_neurobench_2024}).
        Each point corresponds to a network with a different random initialization; colors indicate the different sessions.
    }
    \label{fig:results}
\end{figure}

We first assessed the decoding performance of tinyRSNN and bigRSNN and found that both models provided good estimates of finger velocity (Fig.~\ref{fig:results}).
In particular, both models outperformed previously published feed-forward \ac{SNN} decoders (see Table~\ref{tab:r2_avg}), as well as conventional \acp{ANN} trained on the same dataset (average \ac{R²} values: \textit{ANN}:~0.579, \textit{ANN\_Flat}:~0.615~\citep{yik_neurobench_2024}).
As anticipated, the less constrained bigRSNN exhibited superior performance compared to tinyRSNN across all sessions (Table~\ref{tab:r2_avg}, Fig.~\ref{fig:results}B).
These results suggest that the incorporation of recurrent connections enhances the processing capabilities of \acp{SNN}, conferring advantages in \ac{CST} decoding tasks.

\begin{table}[htb]
    \caption{Session Specific \ac{R²} Scores Compared to Baseline Models.}
    \centering
    \label{tab:r2_avg}
    \begin{tabularx}{\linewidth}{
            >{\hsize=0.12\hsize\arraybackslash}X
            >{\hsize=0.3\hsize\centering\arraybackslash}X
            >{\hsize=0.3\hsize\centering\arraybackslash}X
            >{\hsize=0.45\hsize\centering\arraybackslash}X
            >{\hsize=0.45\hsize\centering\arraybackslash}X
        }
        \toprule
        \multirow{2}{*}{\makebox[\ht\strutbox][c]{\rotatebox{90}{Session}}} & \multicolumn{2}{c}{\textbf{Baseline \citep{yik_neurobench_2024}}} & \multicolumn{2}{c}{\textbf{Ours}}                                                                                     \\
        \cmidrule(lr){2-3}          \cmidrule(lr){4-5}
                                                                            & \multicolumn{1}{c}{\textbf{SNN}}                                  & \multicolumn{1}{c}{\textbf{SNN\_Flat}} & \multicolumn{1}{c}{\textbf{tinyRSNN}} & \multicolumn{1}{c}{\textbf{bigRSNN}} \\
        \midrule
        I1                                                                  & $0.677$                                                           & $0.697$                                & $0.752 \pm 0.003$                     & $0.770 \pm 0.003$                    \\
        I2                                                                  & $0.501$                                                           & $0.577$                                & $0.545 \pm 0.004$                     & $0.585 \pm 0.012$                    \\
        I3                                                                  & $0.599$                                                           & $0.652$                                & $0.746 \pm 0.007$                     & $0.772 \pm 0.006$                    \\
        L1                                                                  & $0.571$                                                           & $0.623$                                & $0.622 \pm 0.003$                     & $0.698 \pm 0.006$                    \\
        L2                                                                  & $0.515$                                                           & $0.568$                                & $0.608 \pm 0.006$                     & $0.629 \pm 0.008$                    \\
        L3                                                                  & $0.620$                                                           & $0.681$                                & $0.690 \pm 0.006$                     & $0.734 \pm 0.005$                    \\
        \cmidrule(lr){1-5}
        \textbf{Mean}                                                       & $\mathbf{0.581}$                                                  & $\mathbf{0.633}$                       & $\mathbf{0.660 \pm 0.002}$            & $\mathbf{0.698 \pm 0.002}$           \\
        \bottomrule
    \end{tabularx}
\end{table}

\begin{table}[htb]
    \centering
    \caption{Computational Cost of tinyRSNN Compared to Resource-Efficient Baseline Models. Bold: Best Values.}
    \begin{tabularx}{\linewidth}{
            >{\hsize=0.04\hsize\arraybackslash}X
            >{\hsize=0.1\hsize\arraybackslash}X
            >{\hsize=0.23\hsize\centering\arraybackslash}X
            >{\hsize=0.23\hsize\centering\arraybackslash}X
            >{\hsize=0.42\hsize\centering\arraybackslash}X
        }
        \toprule
                                                                                    &                                             & \multicolumn{2}{c}{\textbf{Baseline \citep{yik_neurobench_2024}}} & \textbf{Ours}                             \\
        \cmidrule(lr){3-4} \cmidrule(lr){5-5}
        \multicolumn{2}{c}{ }                                                       & \textbf{SNN}                                & \textbf{ANN}                                                      & \textbf{tinyRSNN}                         \\
        \toprule
        \multicolumn{2}{l}{\textbf{Memory footprint [bytes]}}                       & $29248$                                     & $27160$                                                           & $\mathbf{27144\pm0}$                      \\
        \midrule
        \multicolumn{2}{l}{\parbox{3cm}{\textbf{Connection Sparsity}}}              & $0$                                         & $0$                                                               & $\mathbf{0.45\pm0.01}$                    \\
        \midrule
        \multicolumn{2}{l}{\parbox{3cm}{\textbf{Activation Sparsity}}}              & $\mathbf{0.9976}$                           & $0.6755$                                                          & $0.9836\pm0.0001$                         \\
        \midrule
        \multirow{4}{*}{\makebox[\ht\strutbox][l]{\rotatebox{90}{\textbf{SynOps}}}} & \multicolumn{1}{l}{\textbf{Dense}}          & $7300$                                                            & $\mathbf{6237}$        & $13440\pm0 $     \\
        \cmidrule{2-5}
                                                                                    & \multicolumn{1}{l}{\textbf{Effective MACs}} & $\mathbf{0}$                                                      & $4967$                 & $\mathbf{0\pm0}$ \\
        \cmidrule{2-5}
                                                                                    & \multicolumn{1}{l}{\textbf{Effective ACs}}  & $414$                                                             & $\mathbf{0}$           & $304\pm8$        \\
        \midrule
        \multicolumn{2}{l}{\parbox{3cm}{\textbf{Effective Energy Ratio}}}           & $1.36$                                      & $506.50$                                                          & $\mathbf{1}$                              \\
        \bottomrule
    \end{tabularx}
    \label{tab:efficiency}
\end{table}

\subsection{Computational complexity of the tinyRSNN model}
Next, we investigated the memory footprint and computational cost of the models.
We found that tinyRSNN had a smaller memory footprint and higher connection sparsity compared to feed-forward resource-optimized \ac{SNN} decoders \citep{yik_neurobench_2024}, despite its superior decoding performance (cf. Table~\ref{tab:efficiency}).
The tinyRSNN model uses fewer \acp{AC} on average, a direct consequence of activity regularization and pruning.
More precisely, it featured an average connection sparsity of 45\% and an average activation sparsity of 98.36\% (Table~\ref{tab:efficiency}; cf. Tables~\ref{tab:sparsity_per_session} and \ref{tab:synops_per_session} for session-specific results).
To estimate energy consumption on optimized hardware, we computed the effective energy ratio \citep{yin_accurate_2021, horowitz_11_2014}.
By this metric, tinyRSNN reduced energy consumption by 26.6\% compared to the \ac{SNN} decoder from \citet{yik_neurobench_2024}.
Furthermore, it requires approximately 500 times less energy than a traditional \ac{ANN} decoder trained on the same task \citep{yik_neurobench_2024} (see Table~\ref{tab:efficiency}).
These results suggest that small \ac{RSNN} models can effectively leverage activity regularization and synaptic pruning strategies to substantially reduce computational costs while maintaining high decoding performance.

\section{Discussion \& Conclusion}
We developed two \ac{RSNN} models for decoding finger velocities from \acp{CST} and evaluated their decoding performance and energy efficiency.
Our bigRSNN model significantly improved decoding performance compared to existing models.
Additionally, we introduced the tinyRSNN model, tailored for fully implanted \ac{BMI} applications.
This model was designed to find a Pareto optimum between decoding accuracy and energy efficiency.
Our work demonstrates that, through the application of suitable training techniques, the computational complexity of \acp{RSNN} can be remarkably reduced.
By combining small layer sizes, activity regularization, and synaptic pruning, we substantially reduced the computational requirements while hardly impacting decoding performance.
Notably, tinyRSNN exhibits considerable performance improvements over existing models with comparable resource requirements.

While we have demonstrated promising performance improvements, several questions remain open.
For instance, the specific roles of individual factors, such as recurrent connections, in driving these improvements remain unclear.
A detailed analysis of their impact is a key area for future work.
Recent work has highlighted synaptic delays as an architectural feature that can enhance \ac{SNN} performance \citep{zhang2020supervised,hammouamri_learning_2023}.
These delays may also allow for reduced memory requirements while being amenable to efficient hardware implementations \citep{dagostino_denram_2024, moro_role_2024}.
In future work, we will explore whether incorporating synaptic delays could further improve \acp{CST} decoding.
Although tinyRSNN constitutes a parsimonious model that should in principle be suitable for a hardware implementation, we have not formally shown its performance gains on physical hardware.
We intend to implement this model on suitable neuromorphic processors in future work, a crucial step toward realizing practical, fully implanted \acp{BMI}.

In conclusion, \acp{RSNN} have proven to be competitive models for real-time \acp{CST} decoding that outperform existing models.
Moreover, we have shown that their computational complexity can be substantially reduced through suitable training strategies, with minimal impact on accuracy.
These advancements bring us closer to the next generation of efficient, high-performance neural decoders for implantable \acp{BMI} applications, with the potential to elevate the standard of patient care.

\section{Acknowledgments}

This work was supported by
STI 2030-Major Projects (2022ZD0208604, 2021ZD0200300),
ZJU Doctoral Graduate Academic Rising Star Development Program (2023059),
EU’s Horizon Europe Research and Innovation Programme (Grant Agreement No.\ 101070374, CONVOLVE) funded through SERI (Ref.\ 1131-52302),
the Swiss National Science Foundation (Grant Number PCEFP3\_202981),
and the Novartis Research Foundation.

\begin{table}[tbh]
    \centering
    \caption{Quantification of Memory Footprint and Sparsity.}
    \begin{tabularx}{\linewidth}{
            >{\hsize=0.13\hsize\arraybackslash}X
            >{\hsize=0.15\hsize\arraybackslash}X
            >{\hsize=0.46 \hsize\centering\arraybackslash}X
            >{\hsize=0.46\hsize\centering\arraybackslash}X
        }
        \toprule
         & \makebox[\ht\strutbox][l]{\textbf{Session}} & \multicolumn{1}{c}{\textbf{tinyRSNN}} & \multicolumn{1}{c}{\textbf{bigRSNN}} \\
        \midrule

        \multirow{3}{*}{\textbf{\rotatebox{90}{\parbox{1.1cm}{Memory footprint [bytes]}}}}
         & I1 - I3                                     & $\num{21000}$                         & $\num{4636752}$                      \\
         & L1 - L3                                     & $\num{33288}$                         & $\num{5029968}$                      \\
        \cmidrule(lr){2-4}
         & \textbf{Mean}                               & $\mathbf{\num{27144}}$                & $\mathbf{\num{4833360}}$             \\
        \midrule

        \multirow{7}{*}{\textbf{\rotatebox{90}{\parbox{1.2cm}{Connection Sparsity}}}}
         & I1                                          & $0.47 \pm 0.02$                       & $0$                                  \\
         & I2                                          & $0.45 \pm 0.03$                       & $0$                                  \\
         & I3                                          & $0.50 \pm 0.00$                       & $0$                                  \\
         & L1                                          & $0.44 \pm 0.04$                       & $0$                                  \\
         & L2                                          & $0.42 \pm 0.02$                       & $0$                                  \\
         & L3                                          & $0.45 \pm 0.03$                       & $0$                                  \\
        \cmidrule(lr){2-4}
         & \textbf{Mean}                               & $\mathbf{0.45 \pm 0.01}$              & $\mathbf{0}$                         \\
        \midrule

        \multirow{7}{*}{\textbf{\rotatebox{90}{\parbox{1.2cm}{Activation Sparsity}}}}
         & I1                                          & $0.9838 \pm 0.0004$                   & $0.9622 \pm 0.0002$                  \\
         & I2                                          & $0.9853 \pm 0.0003$                   & $0.9718 \pm 0.0006$                  \\
         & I3                                          & $0.9842 \pm 0.0002$                   & $0.9721 \pm 0.0002$                  \\
         & L1                                          & $0.9831 \pm 0.0001$                   & $0.9677 \pm 0.0012$                  \\
         & L2                                          & $0.9832 \pm 0.0002$                   & $0.9674 \pm 0.0010$                  \\
         & L3                                          & $0.9820 \pm 0.0003$                   & $0.9686 \pm 0.0014$                  \\
        \cmidrule(lr){2-4}
         & \textbf{Mean}                               & $\mathbf{0.9836 \pm 0.0001}$          & $\mathbf{0.9683 \pm 0.0005}$         \\
        \bottomrule
    \end{tabularx}
    \label{tab:sparsity_per_session}
\end{table}
\begin{table}[tbh]
    \centering
    \caption{Session Specific and Average Synaptic Operations.}
    \begin{tabularx}{\linewidth}{
            >{\hsize=0.05\hsize\arraybackslash}X
            >{\hsize=0.2\hsize\arraybackslash}X
            >{\hsize=0.5 \hsize\centering\arraybackslash}X
            >{\hsize=0.5\hsize\centering\arraybackslash}X
        }
        \toprule
         & \makebox[\ht\strutbox][l]{\textbf{Session}} & \multicolumn{1}{c}{\textbf{tinyRSNN}} & \multicolumn{1}{c}{\textbf{bigRSNN}} \\
        \midrule

        \multirow{4}{*}{\textbf{\rotatebox{90}{Dense}}}
         & I1 - I3                                     & $\num{10368}$                         & $\num{1157120}$                      \\
         & L1 - L3                                     & $\num{16512}$                         & $\num{1255424}$                      \\
        \cmidrule(lr){2-4}
         & \textbf{Mean}                               & $\mathbf{\num{13440}}$                & $\mathbf{\num{1206272}}$             \\
        \midrule

        \multirow{10}{*}{\textbf{\rotatebox{90}{Effective ACs}}}
         & I1                                          & $299\pm14$                            & $\num{48096}\pm254$                  \\
         & I2                                          & $197\pm12$                            & $\num{34837}\pm659$                  \\
         & I3                                          & $143\pm2$                             & $\num{33289}\pm181$                  \\
        \cmidrule(lr){2-4}
         & \parbox{1.6cm}{\textbf{Mean Indy}}          & $\mathbf{213\pm4}$                    & $\mathbf{\num{38741}\pm\num{190}}$   \\

        \cmidrule(lr){2-4}
         & L1                                          & $354\pm28$                            & $\num{44664}\pm\num{1258}$           \\
         & L2                                          & $404\pm20$                            & $\num{45945}\pm\num{1097}$           \\
         & L3                                          & $427\pm28$                            & $\num{45189}\pm\num{1437}$           \\
        \cmidrule(lr){2-4}
         & \parbox{1.6cm}{\textbf{Mean Loco}}          & $\mathbf{395\pm19}$                   & $\mathbf{\num{45266}\pm\num{1056}}$  \\
        \bottomrule
    \end{tabularx}
    \label{tab:synops_per_session}
\end{table}

\printbibliography

@article{Card2024-rv,
  title     = {An accurate and rapidly calibrating speech neuroprosthesis},
  author    = {Card, Nicholas S and Wairagkar, Maitreyee and Iacobacci, Carrina
               and Hou, Xianda and Singer-Clark, Tyler and Willett, Francis R
               and Kunz, Erin M and Fan, Chaofei and Vahdati Nia, Maryam and
               Deo, Darrel R and Srinivasan, Aparna and Choi, Eun Young and
               Glasser, Matthew F and Hochberg, Leigh R and Henderson, Jaimie M
               and Shahlaie, Kiarash and Stavisky, Sergey D and Brandman, David
               M},
  journal   = {The New England journal of medicine},
  publisher = {Massachusetts Medical Society},
  volume    = 391,
  number    = 7,
  pages     = {609--618},
  month     = aug,
  year      = 2024,
  doi       = {10.1056/NEJMoa2314132},
  pmc       = {PMC11328962},
  pmid      = 39141853,
  issn      = {0028-4793,1533-4406},
  language  = {en}
}

@article{chaudhary2016brain,
  title     = {Brain--computer interfaces for communication and rehabilitation},
  author    = {Chaudhary, Ujwal and Birbaumer, Niels and Ramos-Murguialday, Ander},
  journal   = {Nature Reviews Neurology},
  volume    = {12},
  number    = {9},
  pages     = {513--525},
  year      = {2016},
  publisher = {Nature Publishing Group UK London}
}

@article{dagostino_denram_2024,
  title      = {{{DenRAM}}: Neuromorphic Dendritic Architecture with {{RRAM}} for Efficient Temporal Processing with Delays},
  shorttitle = {{{DenRAM}}},
  author     = {D'Agostino, Simone and Moro, Filippo and Torchet, Tristan and Demira{\u g}, Yi{\u g}it and Grenouillet, Laurent and Castellani, Niccol{\`o} and Indiveri, Giacomo and Vianello, Elisa and Payvand, Melika},
  year       = {2024},
  month      = apr,
  journal    = {Nature Communications},
  volume     = {15},
  number     = {1},
  pages      = {3446},
  publisher  = {Nature Publishing Group},
  issn       = {2041-1723},
  doi        = {10.1038/s41467-024-47764-w},
  langid     = {english}
}

@article{donati_neuromorphic_2024,
  title     = {Neuromorphic hardware for somatosensory neuroprostheses},
  volume    = {15},
  copyright = {2024 The Author(s)},
  issn      = {2041-1723},
  doi       = {10.1038/s41467-024-44723-3},
  language  = {en},
  number    = {1},
  journal   = {Nature Communications},
  author    = {Donati, Elisa and Valle, Giacomo},
  month     = jan,
  year      = {2024},
  note      = {Number: 1
               Publisher: Nature Publishing Group},
  pages     = {556}
}

@article{flesher2021brain,
  title     = {A brain-computer interface that evokes tactile sensations improves robotic arm control},
  author    = {Flesher, Sharlene N and Downey, John E and Weiss, Jeffrey M and Hughes, Christopher L and Herrera, Angelica J and Tyler-Kabara, Elizabeth C and Boninger, Michael L and Collinger, Jennifer L and Gaunt, Robert A},
  journal   = {Science},
  volume    = {372},
  number    = {6544},
  pages     = {831--836},
  year      = {2021},
  publisher = {American Association for the Advancement of Science}
}

@misc{funk_rmsprop_2015-1,
  title        = {{{RMSprop}} Loses to {{SMORMS3}} - {{Beware}} the {{Epsilon}}!},
  author       = {Funk, Simon},
  year         = {2015},
  urldate      = {2022-04-20},
  howpublished = {\url{https://sifter.org/simon/journal/20150420.html}}
}

@misc{hammouamri_learning_2023,
  title         = {Learning {{Delays}} in {{Spiking Neural Networks}} Using {{Dilated Convolutions}} with {{Learnable Spacings}}},
  author        = {Hammouamri, Ilyass and {Khalfaoui-Hassani}, Ismail and Masquelier, Timoth{\'e}e},
  year          = {2023},
  month         = jun,
  number        = {arXiv:2306.17670},
  eprint        = {2306.17670},
  primaryclass  = {cs},
  publisher     = {arXiv},
  archiveprefix = {arXiv},
  langid        = {english}
}

@article{hochberg2012reach,
  title     = {Reach and grasp by people with tetraplegia using a neurally controlled robotic arm},
  author    = {Hochberg, Leigh R and Bacher, Daniel and Jarosiewicz, Beata and Masse, Nicolas Y and Simeral, John D and Vogel, Joern and Haddadin, Sami and Liu, Jie and Cash, Sydney S and Van Der Smagt, Patrick and others},
  journal   = {Nature},
  volume    = {485},
  number    = {7398},
  pages     = {372--375},
  year      = {2012},
  publisher = {Nature Publishing Group UK London}
}

@inproceedings{horowitz_11_2014,
  title     = {1.1 {{Computing}}'s Energy Problem (and What We Can Do about It)},
  booktitle = {2014 {{IEEE International Solid-State Circuits Conference Digest}} of {{Technical Papers}} ({{ISSCC}})},
  author    = {Horowitz, Mark},
  year      = {2014},
  month     = feb,
  pages     = {10--14},
  issn      = {2376-8606},
  doi       = {10.1109/ISSCC.2014.6757323}
}

@article{indiveri_neuromorphic_2011,
  title     = {Neuromorphic {{Silicon Neuron Circuits}}},
  author    = {Indiveri, Giacomo and {Linares-Barranco}, Bernabe and Hamilton, Tara J. and {van Schaik}, Andr{\'e} and {Etienne-Cummings}, Ralph and Delbruck, Tobi and Liu, Shih-Chii and Dudek, Piotr and H{\"a}fliger, Philipp and Renaud, Sylvie and Schemmel, Johannes and Cauwenberghs, Gert and Arthur, John and Hynna, Kai and Folowosele, Fopefolu and Sa{\"i}ghi, Sylvain and {Serrano-Gotarredona}, Teresa and Wijekoon, Jayawan and Wang, Yingxue and Boahen, Kwabena},
  year      = {2011},
  month     = may,
  journal   = {Frontiers in Neuroscience},
  volume    = {5},
  publisher = {Frontiers},
  issn      = {1662-453X},
  doi       = {10.3389/fnins.2011.00073},
  langid    = {english}
}

@inproceedings{liao2022energy,
  title        = {An energy-efficient spiking neural network for finger velocity decoding for implantable brain-machine interface},
  author       = {Liao, Jiawei and Widmer, Lars and Wang, Xiaying and Di Mauro, Alfio and Nason-Tomaszewski, Samuel R and Chestek, Cynthia A and Benini, Luca and Jang, Taekwang},
  booktitle    = {2022 IEEE 4th International Conference on Artificial Intelligence Circuits and Systems (AICAS)},
  pages        = {134--137},
  year         = {2022},
  organization = {IEEE}
}

@misc{moro_role_2024,
  title         = {The {{Role}} of {{Temporal Hierarchy}} in {{Spiking Neural Networks}}},
  author        = {Moro, Filippo and Aceituno, Pau Vilimelis and Kriener, Laura and Payvand, Melika},
  year          = {2024},
  month         = jul,
  number        = {arXiv:2407.18838},
  eprint        = {2407.18838},
  primaryclass  = {cs},
  publisher     = {arXiv},
  archiveprefix = {arXiv},
  langid        = {english}
}

@article{moses_neuroprosthesis_2021,
  title     = {Neuroprosthesis for {{Decoding Speech}} in a {{Paralyzed Person}} with {{Anarthria}}},
  author    = {Moses, David A. and Metzger, Sean L. and Liu, Jessie R. and Anumanchipalli, Gopala K. and Makin, Joseph G. and Sun, Pengfei F. and Chartier, Josh and Dougherty, Maximilian E. and Liu, Patricia M. and Abrams, Gary M. and {Tu-Chan}, Adelyn and Ganguly, Karunesh and Chang, Edward F.},
  year      = {2021},
  month     = jul,
  journal   = {New England Journal of Medicine},
  volume    = {385},
  number    = {3},
  pages     = {217--227},
  publisher = {Massachusetts Medical Society},
  issn      = {0028-4793},
  doi       = {10.1056/NEJMoa2027540}
}

@article{neftci_surrogate_2019,
  title      = {Surrogate {Gradient} {Learning} in {Spiking} {Neural} {Networks}: {Bringing} the {Power} of {Gradient}-based optimization to spiking neural networks},
  volume     = {36},
  issn       = {1053-5888},
  shorttitle = {Surrogate {Gradient} {Learning} in {Spiking} {Neural} {Networks}},
  doi        = {10.1109/MSP.2019.2931595},
  number     = {6},
  journal    = {IEEE Signal Processing Magazine},
  author     = {Neftci, Emre O. and Mostafa, Hesham and Zenke, Friedemann},
  month      = nov,
  year       = {2019},
  pages      = {51--63}
}

@misc{odoherty_nonhuman_2017,
  title     = {Nonhuman {{Primate Reaching}} with {{Multichannel Sensorimotor Cortex Electrophysiology}}},
  author    = {O'Doherty, Joseph E. and Cardoso, Mariana M. B. and Makin, Joseph G. and Sabes, Philip N.},
  year      = {2017},
  month     = may,
  publisher = {Zenodo},
  doi       = {10.5281/zenodo.583331}
}

@article{pandarinath2017high,
  title     = {High performance communication by people with paralysis using an intracortical brain-computer interface},
  author    = {Pandarinath, Chethan and Nuyujukian, Paul and Blabe, Christine H and Sorice, Brittany L and Saab, Jad and Willett, Francis R and Hochberg, Leigh R and Shenoy, Krishna V and Henderson, Jaimie M},
  journal   = {elife},
  volume    = {6},
  pages     = {e18554},
  year      = {2017},
  publisher = {eLife Sciences Publications, Ltd}
}

@inproceedings{paszke_pytorch_2019,
  author    = {Paszke, Adam and Gross, Sam and Massa, Francisco and Lerer, Adam and Bradbury, James and Chanan, Gregory and Killeen, Trevor and Lin, Zeming and Gimelshein, Natalia and Antiga, Luca and Desmaison, Alban and Kopf, Andreas and Yang, Edward and DeVito, Zachary and Raison, Martin and Tejani, Alykhan and Chilamkurthy, Sasank and Steiner, Benoit and Fang, Lu and Bai, Junjie and Chintala, Soumith},
  editor    = {Wallach, H. and Larochelle, H. and Beygelzimer, A. and d'Alché-Buc, F. and Fox, E. and Garnett, R.},
  publisher = {Curran Associates, Inc.},
  booktitle = {Advances in Neural Information Processing Systems 32},
  date      = {2019},
  pages     = {8024--8035},
  title     = {PyTorch: An Imperative Style, High-Performance Deep Learning Library}
}

@article{perez-nieves_neural_2021,
  title   = {Neural Heterogeneity Promotes Robust Learning},
  author  = {{Perez-Nieves}, Nicolas and Leung, Vincent C. H. and Dragotti, Pier Luigi and Goodman, Dan F. M.},
  year    = {2021},
  month   = dec,
  journal = {Nature Communications},
  volume  = {12},
  number  = {1},
  pages   = {5791},
  issn    = {2041-1723},
  doi     = {10.1038/s41467-021-26022-3}
}

@inbook{Rokach2005,
  author    = {Rokach, Lior},
  editor    = {Maimon, Oded
               and Rokach, Lior},
  title     = {Ensemble Methods for Classifiers},
  booktitle = {Data Mining and Knowledge Discovery Handbook},
  year      = {2005},
  publisher = {Springer US},
  address   = {Boston, MA},
  pages     = {957--980},
  isbn      = {978-0-387-25465-4},
  doi       = {10.1007/0-387-25465-X_45}
}

@article{rossbroich_fluctuation-driven_2022,
  title   = {Fluctuation-Driven Initialization for Spiking Neural Network Training},
  author  = {Rossbroich, Julian and Gygax, Julia and Zenke, Friedemann},
  year    = {2022},
  month   = dec,
  journal = {Neuromorphic Computing and Engineering},
  volume  = {2},
  number  = {4},
  pages   = {044016},
  issn    = {2634-4386},
  doi     = {10.1088/2634-4386/ac97bb},
  langid  = {english}
}

@article{taeckens_spiking_2024,
  title     = {A Spiking Neural Network with Continuous Local Learning for Robust Online Brain Machine Interface},
  author    = {Taeckens, Elijah A. and Shah, Sahil},
  year      = {2024},
  month     = jan,
  journal   = {Journal of Neural Engineering},
  volume    = {20},
  number    = {6},
  pages     = {066042},
  publisher = {IOP Publishing},
  issn      = {1741-2552},
  doi       = {10.1088/1741-2552/ad1787},
  langid    = {english}
}

@article{willett2021high,
  title     = {High-performance brain-to-text communication via handwriting},
  author    = {Willett, Francis R and Avansino, Donald T and Hochberg, Leigh R and Henderson, Jaimie M and Shenoy, Krishna V},
  journal   = {Nature},
  volume    = {593},
  number    = {7858},
  pages     = {249--254},
  year      = {2021},
  publisher = {Nature Publishing Group UK London}
}

@article{wolf2008thermal,
  title     = {Thermal considerations for the design of an implanted cortical brain--machine interface (BMI)},
  author    = {Wolf, Patrick D and Reichert, WM},
  journal   = {Indwelling Neural Implants: Strategies for Contending with the In Vivo Environment},
  pages     = {33--38},
  year      = {2008},
  publisher = {CRC Press/Taylor \& Francis Boca Raton}
}

@article{yik_neurobench_2024,
  title      = {{{NeuroBench}}: {{A Framework}} for {{Benchmarking Neuromorphic Computing Algorithms}} and {{Systems}}},
  shorttitle = {{{NeuroBench}}},
  author     = {Yik, Jason and {Van den Berghe}, Korneel and {den Blanken}, Douwe and Bouhadjar, Younes and Fabre, Maxime and Hueber, Paul and Kleyko, Denis and {Pacik-Nelson}, Noah and Sun, Pao-Sheng Vincent and Tang, Guangzhi and Wang, Shenqi and Zhou, Biyan and Hasan Ahmed, Soikat and Vathakkattil Joseph, George and Leto, Benedetto and Micheli, Aurora and Mishra, Anurag Kumar and Lenz, Gregor and Sun, Tao and Ahmed, Zergham and Akl, Mahmoud and Anderson, Brian and Andreou, Andreas G. and Bartolozzi, Chiara and Basu, Arindam and Bogdan, Petrut and Bohte, Sander and Buckley, Sonia and Cauwenberghs, Gert and Chicca, Elisabetta and Corradi, Federico and {de Croon}, Guido and Danielescu, Andreea and Daram, Anurag and Davies, Mike and Demirag, Yigit and Eshraghian, Jason and Fischer, Tobias and Forest, Jeremy and Fra, Vittorio and Furber, Steve and Furlong, P. Michael and Gilpin, William and Gilra, Aditya and Gonzalez, Hector A. and Indiveri, Giacomo and Joshi, Siddharth and Karia, Vedant and Khacef, Lyes and Knight, James C. and Kriener, Laura and Kubendran, Rajkumar and Kudithipudi, Dhireesha and Liu, Yao-Hong and Liu, Shih-Chii and Ma, Haoyuan and Manohar, Rajit and {Margarit-Taul{\'e}}, Josep Maria and Mayr, Christian and Michmizos, Konstantinos and Muir, Dylan and Neftci, Emre and Nowotny, Thomas and Ottati, Fabrizio and Ozcelikkale, Ayca and Panda, Priyadarshini and Park, Jongkil and Payvand, Melika and Pehle, Christian and Petrovici, Mihai A. and Pierro, Alessandro and Posch, Christoph and Renner, Alpha and Sandamirskaya, Yulia and Schaefer, Clemens J.S. and {van Schaik}, Andr{\'e} and Schemmel, Johannes and Schmidgall, Samuel and Schuman, Catherine and Seo, Jae-sun and Sheik, Sadique and Bam Shrestha, Sumit and Sifalakis, Manolis and Sironi, Amos and Stewart, Matthew and Stewart, Kenneth and Stewart, Terrence C. and Stratmann, Philipp and Timcheck, Jonathan and T{\"o}men, Nergis and Urgese, Gianvito and Verhelst, Marian and Vineyard, Craig M. and Vogginger, Bernhard and Yousefzadeh, Amirreza and Tuz Zohora, Fatima and Frenkel, Charlotte and Janapa Reddi, Vijay},
  year       = {2024},
  month      = jan,
  journal    = {NeuroBench},
  volume     = {2304.04640},
  publisher  = {arXiv.org},
  doi        = {10.48550/arXiv.2304.04640}
}

@article{yin_accurate_2021,
  title     = {Accurate and Efficient Time-Domain Classification with Adaptive Spiking Recurrent Neural Networks},
  author    = {Yin, Bojian and Corradi, Federico and Boht{\'e}, Sander M.},
  year      = {2021},
  month     = oct,
  journal   = {Nature Machine Intelligence},
  volume    = {3},
  number    = {10},
  pages     = {905--913},
  publisher = {Nature Publishing Group},
  issn      = {2522-5839},
  doi       = {10.1038/s42256-021-00397-w},
  copyright = {2021 The Author(s), under exclusive licence to Springer Nature Limited},
  langid    = {english}
}

@article{zhang2020supervised,
  title     = {Supervised learning in spiking neural networks with synaptic delay-weight plasticity},
  author    = {Zhang, Malu and Wu, Jibin and Belatreche, Ammar and Pan, Zihan and Xie, Xiurui and Chua, Yansong and Li, Guoqi and Qu, Hong and Li, Haizhou},
  journal   = {Neurocomputing},
  volume    = {409},
  pages     = {103--118},
  year      = {2020},
  publisher = {Elsevier}
}

@data{zhou_ieee_2024,
  title     = {{{IEEE BioCAS}} 2024 {{Grand Challenge}} on {{Neural Decoding}} for {{Motor Control}} of Non-{{Human Primates}}},
  author    = {Zhou, Biyan and Sun, Pao-Sheng Vincent and Yik, Jason and Frenkel, Charlotte and Reddi, Vijay Janapa and O'Doherty, Joseph E. and Cardoso, Mariana M. B. and Makin, Joseph G. and Sabes, Philip N. and Basu, Arindam},
  year      = {2024},
  month     = may,
  day       = 31,
  publisher = {IEEE Dataport},
  doi       = {10.21227/bp1f-te92},
  date      = {2024-05-31}
}

\end{document}